\documentclass[prd,aps,nofootinbib,onecolumn,showpacs,12pt]{revtex4-1}
\usepackage{graphicx,epsfig}

\usepackage{epsf}
\usepackage{graphicx}

\begin{document}

\title{  \bf   Thermal QCD Sum Rules Study of Vector Charmonium and Bottomonium States}
\author{ E. Veli Veliev$^{*1}$,
K. Azizi$^{\dag2}$, H. Sundu$^{*3}$, G. Kaya$^{*4}$, A. T\"urkan$^{*5}$ \\
$^{*}$Department of Physics, Kocaeli University, 41380 Izmit,
Turkey\\
 $^{\dag}$Physics Division,  Faculty of Arts and Sciences,
Do\u gu\c s University,
 Ac{\i}badem-Kad{\i}k\"oy, \\ 34722 Istanbul, Turkey\\
 $^1$ e-mail:elsen@kocaeli.edu.tr\\
$^2$e-mail:kazizi@dogus.edu.tr\\
$^3$email:hayriye.sundu@kocaeli.edu.tr\\
$^4$email:gulsahbozkir@kocaeli.edu.tr\\
$^5$email:arzu.turkan1@kocaeli.edu.tr }

\begin{abstract}
We calculate the masses and leptonic decay constants of the heavy
vector quarkonia, $J/\psi$ and  $\Upsilon$  mesons at finite temperature. In particular, considering the thermal spectral density as well as additional operators
coming up at finite temperature, the thermal QCD sum rules are
acquired. Our numerical calculations demonstrate   that the masses and decay
constants are insensitive to the variation of temperature up to $T\cong 100 ~MeV$, however after this
point, they start to fall  altering the temperature.
At deconfinement temperature, the decay constants attain
roughly to 45\% of their vacuum values, while the masses are
diminished about 12\%, and 2.5\%  for $J/\psi$ and $\Upsilon$ states,
respectively. The obtained results at zero temperature are in  good
consistency with the existing experimental data as well as
predictions of the other nonperturbative models. Considerable decreasing in the values of the decay constants can be considered as a sign of the quark gluon plasma  phase transition.
\end{abstract}

\maketitle


\section{Introduction }
Investigation of the heavy mesons can play essential role in understanding the vacuum properties of the nonperturbative QCD \cite{Shifman}. In particular, analysis of the variation of the parameters of the heavy quarkonia,
 namely bottonium ($\bar b b$) and charmonium ($\bar c c$) in hadronic medium  with respect to the temperature can give valuable information about the QCD vacuum and transition to the quark gluon plasma (QGP) phase.
Determination of the hadronic properties of the vector mesons in hot and dense QCD medium has become one of the most important research subject in the last twenty years both theoretically and experimentally. 
$J/\psi$ suppression effect due to color screening can be considered as an important evidence for QGP \cite{Matsui and Satz}. This suppression effect has been observed experimentally in heavy ion collision experiments
 held in super proton synchrotron (SPS) at CERN  and relativistic heavy ion collider (RHIC) at BNL.

Properties of the heavy mesons in vacuum have been probed widely in the literature using the nonperturbative approaches like QCD sum rules, nonrelativistic potential models, lattice theory, heavy quark effective theory and chiral perturbation
 theory. However, in expansion of most of these models to finite temperature we are face to face with some difficulties. QCD sum rules which is based on the operator product expansion (OPE), QCD Lagrangian and quark-hadron
 duality, is one of the most informative, applicable and predictive models in hadron physics \cite{L.J.Reinders,P. Colangelo}. The thermal version of this model proposed by Bochkarev  and
Shaposhnikov \cite{Bochkarev} has some new features at $T\neq0$ \cite{E.V. Shuryak,T.
Hatsuda,S. Mallik}. One of the new feature is the interaction of the particles existing in the medium with the currents which demands the modification of the hadronic spectral function.
 The other new picture of the thermal QCD  is breakdown of the  Lorentz invariance via the choice
of reference frame. Due to residual O(3) symmetry at finite temperature, more operators with the same dimensions
appear in the OPE comparing to the QCD sum rules in vacuum. Thermal version of QCD sum rules has been successfully used
to study the thermal properties of light \cite{S.Mallik
Mukherjee,S.Mallik sarkar,E.V.Veliev},
heavy-light\cite{C.A.Dominguez,C.A.Dominguez2,E.V.Veliev2} and
heavy-heavy \cite{F. Klingl,K.Morita,K.Morita2,E.V.Veliev3} mesons
as a trusty and well-established approach.

In the present work, we calculate the mases and decay constants of the heavy vector quarkonia $J/\psi$ ($\bar c  c$) and  $\Upsilon$ ($\bar b  b$) in the framework of the thermal QCD sum rules.
Using the thermal quark propagator, we calculate the expression for the spectral density in one loop approach. Taking into account also the two loop perturbative contributions in $\alpha_s$ order  \cite{Shifman,L.J.Reinders}
as well as new nonperturbative contributions arising in thermal QCD in addition to vacuum version, we acquire  thermal QCD sum rules for the masses and decay constants. Using the results of the energy density
 for the interval $T= (0-170)~MeV$ obtained via Chiral perturbation theory  \cite{P.Gerber} as well as the values of the energy density and gluon condensates obtained in the region $T= (100-170)~MeV$ 
via lattice QCD \cite{M.Cheng,D.E.Miller}, we present the sensitivity of the masses and decay constants of the $J/\psi$ and $\Upsilon$ heavy vector mesons on the temperature. In our calculations, we also use 
the temperature dependent two loop  expression for the strong coupling constant obtained using the perturbation theory and improved by the lattice results \cite{K.Morita,O.Kaczmarek}. We see that the values of the
decay constants  decrease considerably near to the critical or deconfinement temperature comparing to their values in vacuum. This can be considered as a sign of the QGP phase transition.

The rest of the paper is organized as follows. In the next two sections, we derive  thermal QCD sum rules for the considered observables. The last section is devoted to the numerical analysis of the observables and present
their temperature dependency as well as our discussion.
\section{OPE of Thermal Correlation Function for Heavy-heavy Vector Mesons }

To obtain the thermal QCD sum rules for physical quantities, we need to calculate the convenient thermal correlation function in two different ways: in terms of QCD degrees of freedom and in terms of hadronic parameters.
 In QCD side, the correlation function is evaluated via  OPE which helps us expand the time ordering product of currents in terms of
operators with different dimensions. In the present section, we obtain the OPE for the considered quantities.  We begin by considering the following two point thermal correlation
function:
\begin{eqnarray}\label{Eq1}
\Pi_{\mu\nu}\Big(q,T\Big)=i \int d^{4}x~ e^{iq\cdot x} Tr\Big(\rho~
{\cal T}\Big(J_{\mu}(x)J^{\dag}_{\nu}(0)\Big)\Big),
\end{eqnarray}
where $J_{\mu}(x)=:\overline{Q}(x)\gamma_{\mu} Q(x):$ with $Q=b$ or $c$ is the vector
current, ${\cal T}$ indicates the time ordered product and
$\rho=e^{-\beta H}/Tr e^{-\beta H}$ is the thermal density matrix of
QCD at temperature $T=1/\beta$. As we previously mentioned,  the Lorentz invariance breaks down via the choice of reference frame at which the matter is at rest. However, using the four velocity vector $u_{\mu}$ 
of the matter, we can define Lorentz invariant quantities such as $\omega=u\cdot q$ and
$\overline{q}^{2}=\omega^2-q^2$. By the help of these quantities, the aforesaid thermal correlation function can be expressed in terms of two independent tensors
$P_{\mu\nu}$ and 
$Q_{\mu\nu}$ at finite temperature \cite{S.Mallik
Mukherjee}, i.e.,
\begin{eqnarray}\label{Eq2}
\Pi_{\mu\nu}\Big(q,T\Big)=Q_{\mu\nu} \Pi_{l}(q^2,\omega)+P_{\mu\nu}
\Pi_{t}(q^2,\omega), 
\end{eqnarray}
where
\begin{eqnarray}
 P_{\mu\nu}&=&-g_{\mu\nu}+\frac{q_{\mu}q_{\nu}}{q^2}-\frac{q^2}
{\overline{q}^2}\widetilde{u}_{\mu}\widetilde{u}_{\nu},\nonumber\\
Q_{\mu\nu}&=&\frac{q^4}{\overline{q}^2}\widetilde{u}_{\mu}\widetilde{u}_{\nu},
\end{eqnarray}
and  $\widetilde{u}_{\mu}=u_{\mu}-\omega~q_{\mu}/q^2$. The functions $\Pi_{l}$ and  $\Pi_{t}$ are the following Lorentz invariant functions:
\begin{eqnarray}\label{Eq3}
\Pi_{l}\Big(q^2,\omega
\Big)=\frac{1}{\overline{q}^2}u^{\mu}\Pi_{\mu\nu}u^{\nu},
\end{eqnarray}
\begin{eqnarray}\label{Eq31}
\Pi_{t}\Big(q^2,\omega
\Big)=-\frac{1}{2}\Big(g^{\mu\nu}\Pi_{\mu\nu}+\frac{q^2}{\overline{q}^2}u^{\mu}\Pi_{\mu\nu}u^{\nu}\Big).
\end{eqnarray}

It can be shown that in the limit $|\textbf{q}|\rightarrow 0$, the $\Pi_{t}$ function can be expressed as $\Pi_{t}=-\frac{1}{3}g^{\mu\nu}\Pi_{\mu\nu}$  \cite{S.Mallik sarkar} and one can easily find 
the $\Pi_{t}\Big(q_0,|\textbf{q}|=0\Big)=q_0^2~\Pi_{l}\Big(q_0,|\textbf{q}|=0\Big)$ relation between two $\Pi_{l}$ and $\Pi_{t}$ functions. In real time thermal field theory,  the function
$\Pi_{l}(q^2,\omega)$ or $\Pi_{t}(q^2,\omega)$  can be written in $2\times
2$ matrix form and elements of this matrix depend on only one analytic function
 \cite{R.L.Kobes}. Therefore, 
calculation of the  11-component of this matrix is sufficient to
determine completely the dynamics of the corresponding two-point
function. It can also be shown that in the fixed value of   $|\textbf{q}|$, the spectral representation of the thermal correlation function can be written as  \cite{S.Mallik Mukherjee}:
\begin{eqnarray}\label{Eq4}
\Pi_{l,t}\Big(q_0^2,
T\Big)=\int^{\infty}_{0}{dq_0^{\prime}}^{2}~\frac{\rho_{l,t}\Big({q_0^{\prime}}^{2},
T\Big)}{{q_0^{\prime}}^{2}+Q_0^2},
\end{eqnarray}
 where $Q_0^2=-q_0^2$, and
\begin{eqnarray}
\rho_{l,t}\Big(q_0^{2},
T\Big)=\frac{1}{\pi}Im\Pi_{l,t}\Big(q_0^{2}, T\Big)\tanh\frac{\beta
q_0}{2}.
\end{eqnarray}

The thermal correlation function of Eq. (\ref{Eq1})
can be written  in momentum space as:
\begin{eqnarray}\label{Eq5}
\Pi_{\mu\nu}\Big(q, T\Big)=i\int
\frac{d^4k}{(2\pi)^4}Tr\Big[\gamma_{\mu}S(k)\gamma_{\nu}S(k-q)\Big],
\end{eqnarray}
where, we consider the 11-component of the $S(k)$  (thermal quark propagator) which
is expressed as a sum of its vacuum expression and a term depending
on the Fermi distribution function \cite{A.Das}
\begin{eqnarray}\label{Eq6}
S(k)=(\gamma^{\mu}k_{\mu}+m)\Big(\frac{1}{k^2-m^2+i\varepsilon}+2\pi
in(|k_0|)\delta(k^2-m^2)\Big),
\end{eqnarray}
where  $n(x)=[\exp(\beta
x)+1]^{-1}$ is the Fermi distribution function.
Now, we insert the propagator of Eq. (\ref{Eq6}) in Eq. (\ref{Eq5}) and
consider $\Pi_{1}(q,T)=g^{\mu\nu}\Pi_{\mu\nu}(q,T)$ function. Carrying out the integral over $k_0$, we obtain the imaginary part of the 
$\Pi_{1}(q,T)$ in the following form:
\begin{eqnarray}\label{Eq7}
Im\Pi_{1}(q,T)=L(q_0)+L(-q_0),
\end{eqnarray}
where
\begin{eqnarray}\label{Lq0}
L(q_0)&=&N_c\int
\frac{d\textbf{k}}{4\pi^2}\frac{\omega_1^2-\textbf{k}^2+\textbf{k}\cdot
\textbf{q}-\omega_1 q_0-2m^2}{\omega_1 \omega_2}
\nonumber\\
&\times&\Big([(1-n_1)(1-n_2)+n_1n_2]\delta(q_0-\omega_1-\omega_2)-[(1-n_1)n_2+(1-n_2)n_1]
\delta(q_0-\omega_1+\omega_2)\Big),\nonumber\\
\end{eqnarray}
and $n_1=n(\omega_1)$, $n_2=n(\omega_2)$,
$\omega_1=\sqrt{\textbf{k}^2+m^2}$ and
$\omega_2=\sqrt{(\textbf{k}-\textbf{q})^2+m^2}$. The terms without the Fermi distribution functions show the vacuum
contributions but those including the Fermi distribution functions depict medium
contributions. The delta-functions in the different terms of
Eq. (\ref{Lq0}) control the regions of non-vanishing imaginary parts
of $\Pi_1(q,T)$, which define the position of  branch cuts
\cite{Bochkarev}. After straightforward calculations, the
annihilation and scattering parts of $\rho_{1}\Big(q_0^{2},
T\Big)=\frac{1}{\pi}Im\Pi_{1}\Big(q_0^{2}, T\Big)\tanh\frac{\beta
q_0}{2}$ at nonzero momentum can be written as:
\begin{eqnarray}\label{Eq9}
\rho_{1,a}=\frac{-3q^2}{8\pi^2}(3-\nu^2)\Big[\nu-\int^{\nu}_{-\nu}dx~
n_+(x)\Big]~~~~~~~~~\mbox{for}~~~~~~ 4m^2+\textbf{q}^2\leq
q_0^2\leq\infty,
\end{eqnarray}
\begin{eqnarray}\label{Eq91}
\rho_{1,s}=\frac{3q^2}{16\pi^2}(3-\nu^2)\int^{\infty}_{\nu}dx
\Big[n_-(x)-n_+(x)\Big]~~~~~~\mbox{for}~~~~~~~q_0^2\leq~\textbf{q}^2,
\end{eqnarray}
where $\nu(q_0^2)=\sqrt{1-4m^2/q_0^2}$,
$n_+(x)=n\Big[\frac{1}{2}(q_0+|\textbf{q}|x)\Big]$ and
$n_-(x)=n\Big[\frac{1}{2}(|\textbf{q}|x-q_0)\Big]$.
From the similar manner, one can calculate also the function $\Pi_{2}(q,T)=u^{\mu}\Pi_{\mu\nu}(q,T)u^{\nu}$. Using the obtained results in  Eqs. (\ref{Eq3}) and
(\ref{Eq31}),  the annihilation and
scattering parts of $\rho_t$  at nonzero momentum is obtained as:
\begin{eqnarray}\label{Eq10}
\rho_{t,a}&=&\frac{3q^2}{32\pi^2}\int^{\nu}_{-\nu}dx(2-\nu^2+x^2)[1-2n_+(x)],
\end{eqnarray}
\begin{eqnarray}\label{Eq11}
\rho_{t,s}&=&-\frac{3q^2}{32\pi^2}\int^{\infty}_{\nu}dx(2-\nu^2+x^2)[n_-(x)-n_+(x)].
\end{eqnarray}
The annihilation part of $\rho_{l}$, i.e., $\rho_{l,a}$ and its scattering part $\rho_{l,s}$ also  at nonzero
momentum can be found from
Eqs. (\ref{Eq10}) and (\ref{Eq11}) replacing the coefficient $(2-\nu^2+x^2)$ by
$2(1-x^2)$.

In our calculations, we also take into account 
the  perturbative two-loop order $\alpha_{s}$
correction to the spectral density.  This correction at zero temperature can be
written as \cite{Shifman,L.J.Reinders}:
\begin{eqnarray}\label{Pi1}
\rho_{\alpha_{s}}(s)=\alpha_{s} \frac{s}{6\pi^2}\nu (s)\Big(3-\nu^{2}(s)\Big)
\Big[\frac{\pi}{2 \nu (s)}-\frac{1}{4}\Big(3+\nu (s)\Big)
\Big(\frac{\pi}{2}-\frac{3}{4\pi}\Big)\Big],
\end{eqnarray}
where, we replace the strong coupling $\alpha_{s}$ in Eq. (\ref{Pi1}) with its temperature dependent lattice improved expression
$\alpha(T)=2.095(82) \frac{g^2 (T)} {4\pi}$
\cite{K.Morita,O.Kaczmarek}, here

\begin{eqnarray}\label{geks2T}
g^{-2}(T)=\frac{11}{8\pi^2}\ln\Big(\frac{2\pi
T}{\Lambda_{\overline{MS}}}\Big)+\frac{51}{88\pi^2}\ln\Big[2\ln\Big(\frac{2\pi
T}{\Lambda_{\overline{MS}}}\Big)\Big].
\end{eqnarray}

where $\Lambda_{\overline{MS}}=T_c/1.14(4)$ and $T_c=0.160GeV$.

Now, we proceed to calculate  the  nonperturbative part in QCD side. For this aim, we use
the nonperturbative part of the quark propagator in an external
gluon field, $A^a_{\mu}(x)$ in the Fock-Schwinger gauge,
$x^{\mu}A^a_{\mu}(x)=0 $. Taking into account one  and two gluon
lines attached to the quark line, the massive quark propagator  can be written in
momentum space as \cite{L.J.Reinders}:
\begin{eqnarray}\label{Saap}
S^{aa^{\prime}nonpert}(k)&=&
-\frac{i}{4}g (t^{c})^{aa^{\prime}}
G^{c}_{\kappa\lambda}(0)\frac{1}{(k^2-m^2)^2}\Big[\sigma_{\kappa\lambda}
(\not\!k+m)+(\not\!k+m)\sigma_{\kappa\lambda}\Big]
\nonumber\\
&-&\frac{i}{4} g^2
(t^{c}t^{d})^{aa^{\prime}}G^{c}_{\alpha\beta}(0)G^{d}_{\mu\nu}(0)
\frac{\not\!k+m}{(k^2-m^2)^5}(f_{\alpha\beta\mu\nu}+f_{\alpha\mu\beta\nu}
+f_{\alpha\mu\nu\beta})(\not\!k+m),\nonumber\\
\end{eqnarray}
where,
\begin{eqnarray}\label{f}
f_{\alpha\beta\mu\nu}=\gamma_{\alpha}(\not\!k+m)\gamma_{\beta}(\not\!k+m)
\gamma_{\mu}(\not\!k+m)\gamma_{\nu}.
\end{eqnarray}
To go on, we also  need to know the expectation
value $\langle Tr G_{\alpha\beta}G_{\mu\nu}\rangle$. The Lorentz
covariance at finite temperature permits us to write the general
structure of this expectation value in the following manner:
\begin{eqnarray}\label{TrGG} \langle Tr^c G_{\alpha \beta} G_{\mu \nu}
\rangle &=& \frac{1}{24} (g_{\alpha \mu} g_{\beta \nu} -g_{\alpha
\nu} g_{\beta \mu})\langle G^a_{\lambda \sigma} G^{a \lambda \sigma}\rangle \nonumber \\
 &+&\frac{1}{6}\Big[g_{\alpha \mu}
g_{\beta \nu} -g_{\alpha \nu} g_{\beta \mu}-2(u_{\alpha} u_{\mu}
g_{\beta \nu} -u_{\alpha} u_{\nu} g_{\beta \mu} -u_{\beta} u_{\mu}
g_{\alpha \nu} +u_{\beta} u_{\nu} g_{\alpha \mu})\Big]\langle
u^{\lambda} {\Theta}^g _{\lambda \sigma} u^{\sigma}\rangle, \nonumber\\
\end{eqnarray}
where,   $u^{\mu}$ as we also previously mentioned is  the four-velocity of the heat bath and it is
introduced to restore Lorentz invariance formally in the thermal
field theory. In the rest  frame of the heat bath  $ u^{\mu} = (1,
0, 0, 0)$ and $u^2 = 1$. Furthermore, $\Theta^{g}_{\lambda \sigma}$ is   the
traceless gluonic part of the stress-tensor of the QCD.
 Up to terms required for our calculations, the non perturbative part of  massive
quark propagator at finite temperature is obtained as:
\begin{eqnarray}\label{Saap1}
S^{aa^{\prime }nonpert}(k)&=& -\frac{i}{4}g (t^{c})^{aa^{\prime}}
G^{c}_{\kappa\lambda}\frac{1}{(k^2-m^2)^2}\Big[\sigma_{\kappa\lambda}
(\not\!k+m)+(\not\!k+m)\sigma_{\kappa\lambda}\Big]
\nonumber\\
&+&\frac{i~g^2~\delta^{aa^{\prime}}} {3~(k^2-m^2)^4}\Big\{\frac{m
(k^2+m \not\!k)}{4}\langle
G^{c}_{\alpha\beta}G^{c\alpha\beta}\rangle+\frac{1}{3(k^2-m^2) }
\Big[ m (k^2-m^2)\Big(k^2-4(k\cdot u)^2\Big)
\nonumber\\
&+&(m^2-k^2)\Big(-m^2 +4(k\cdot u)^2\Big)\not\!k+4(k\cdot
u)(m^2-k^2)^2\not\!u\Big]\langle
u^{\alpha}\Theta^{g}_{\alpha\beta}u^{\beta}\rangle\Big\}.
\end{eqnarray}
Using the above expression and after straightforward but lengthy
calculations, the nonperturbative part  in QCD side is  obtained
as:
\begin{eqnarray}\label{NonPert}
\Pi_t^{nonpert}&=& \int_{0}^{1}~dx\Big\{-
\frac{\langle\alpha_{s}G^{2}\rangle}{72 \pi \Big[m^2 +q^2
(-1+x)x\Big]^4}\Big[6q^6(-1+x)^4x^4+6m^2q^4x^2(-1+x)^2 (1-6x+6x^2)
\nonumber\\
&+&m^6(5-32x+42x^2-20x^3+10x^4)+m^4q^2x\Big(-14+95x-140x^2
+65^3+6x^4-2x^5\Big)\Big]
\nonumber\\
&-&\frac{\alpha_{s}\langle
u^{\alpha}\Theta^{g}_{\alpha\beta}u^{\beta}\rangle}{54 \pi \Big[m^2
+q^2(-1+x)x\Big]^4}\Big[x(-1+x)
\Big(4q^4x^2(1-3x+2x^2)^2+m^4(12-35x+21x^2+28x^3\nonumber\\&-&14x^4)
+m^2q^2x (-13+55x-82x^2+36x^3+6x^4-2x^5)\Big)\Big(q^2-4(q\cdot
u)^2\Big)\Big] \Big\},
\end{eqnarray}
where, $\langle G^{2}\rangle=\langle
G^{c}_{\alpha\beta}G^{c\alpha\beta}\rangle$.

\section{Phenomenological part and Thermal Sum rules }

Now, we turn our attention to calculate the physical or phenomenological side
of the correlation function. For this aim, we  insert a complete set of physical
intermediate state to  Eq. (\ref{Eq1}) and perform integral over $x$. Isolating the ground state, we get
\begin{eqnarray}\label{phepi}
\Pi_{\mu\nu}(q)=\sum_{\lambda}\frac{{\langle}0|J_{\mu}|V(q,\lambda){\rangle}
{\langle}V(q,\lambda)|J_{\nu}^{\dag}|0{\rangle}}{m_V^2-q^2}+.....,
\end{eqnarray}
where the hadronic states $\{|V(q,\lambda){\rangle}\}$ form a
complete set and $.....$ indicate the contributions of excited
vector mesons and continuum states.

In order to obtain thermal sum rules, now we  equate the spectral
representation and results of operator product expansion for
amplitudes $\Pi_{l}(q^2,\omega)$ or $\Pi_{t}(q^2,\omega)$ at
sufficiently high $Q_0^2$. When performing numerical results, we should exchange our reference to one at which the particle is at rest, i.e., we shall set $|\textbf{q}|\rightarrow 0$. In this limit since the functions
$\Pi_{l}$ and  $\Pi_{t}$ are related to each other, it is enough to use one of them to acquire thermal sum rules. Here, we use the function  $\Pi_{t}$. When we use the standard spectral representation, if the spectral density
 at $s\rightarrow\infty$ does not approach to zero, in this case the correlation function is expressed in terms of a diverge integral. In such a case, to overcome this problem, 
 we subtract first few terms of its
Taylor expansion at $q^2=0$ from $\Pi_{t}(q^2,\omega)$,
\begin{eqnarray}\label{phepit}
\Pi_{t}(q_0^2,|\textbf{q}|)=\Pi_{t}(0)+
\Big(\frac{d\Pi_t}{dQ_0^2}\Big)_{Q_0^2=0} Q_0^2+
\frac{Q_0^4}{\pi}\int^{\infty}_{0}\frac{\rho_t(s)}{s^2(s+Q_0^2)}ds.
\end{eqnarray}

Equating the OPE and hadronic representations of the correlation
function and applying quark-hadron duality,  our
sum-rule takes the form:
\begin{eqnarray}\label{f2Q4}
\frac{f_V^2
Q_0^4}{(m_V^2+Q_0^2)~m_V^2}=Q_0^4\int^{s_0}_{4m^2}\frac{[\rho_{t,a}(s)+\rho_{\alpha_s}(s)]}
{s^2(s+Q_0^2)}ds+\int_{0}^{|\textbf{q}|^2}
\frac{\rho_{t,s}}{s+Q_0^2}ds+\Pi_{t}^{nonpert},
\end{eqnarray}
where,  for simplicity, the total decay width of meson has been neglected. 
The decay constant $f_V$ is defined by the matrix element of the
current $J_{\mu}$ between the vacuum and the vector-meson state, i.e.,
\begin{eqnarray}\label{Jmu}
{\langle}0|J_{\mu}|V(q,\lambda){\rangle}=f_V m_V
\varepsilon^{(\lambda)}_{\mu}.
\end{eqnarray}
In derivation of Eq. (\ref{f2Q4}) we have also used summation over
polarization states,
$\sum_{\lambda}\varepsilon^{(\lambda)^*}_{\mu}\varepsilon^{(\lambda)}_{\nu}=-(g^{\mu\nu}-q_{\mu}q_{\nu}/m_V^2)$.
The Borel transformation removes subtraction terms in the dispersion
relation and also exponentially suppresses the contributions coming from the excited
resonances and continuum states heavier than considered vector ground 
states. Applying Borel transformation with respect to $Q_0^2$ to
both sides of  Eq.(\ref{f2Q4} ), we obtain
\begin{eqnarray}\label{f2m2}
f_V^2m_V^2\exp\Big(-\frac{m_V^2}{M^2}\Big)=\int^{s_0}_{4m^2}ds~[\rho_{t,a}(s)+\rho_{\alpha_s}(s)]e^{-\frac{s}{M^2}}
+\int^{|\textbf{q}|^2}_{0}ds~\rho_{t,s}(s)e^{-\frac{s}{M^2}}+\widehat{B}\Pi^{nonpert}_t
.\nonumber\\
\end{eqnarray}
As we also previously mentioned, when doing numerical analysis, we will set $|\textbf{q}|\rightarrow 0$ representing the rest frame of the  particle. In this case, the scattering cut shrinks to
a point and the spectral density becomes a singular function. Hence, the
second term in the right side of Eq. (\ref{f2m2}) must be detailed
analyzed. Similar analysis has been also performed in \cite{Bochkarev,S.Mallik Mukherjee}. Detailed analysis shows that
\begin{eqnarray}\label{lim}
\lim_{|\textbf{q}|\rightarrow0}\int_{0}^{|\textbf{q}|^2}ds\rho_{t,s}(s)exp\Big(-\frac{s}{M^2}\Big)=0
.
\end{eqnarray}
In Eq. (\ref{f2m2}), $\widehat{B}\Pi^{nonpert}_t$ shows the nonperturbative part of
QCD side in Borel transformed scheme, which is given by:
\begin{eqnarray}\label{BorelNonPert}
&&\hat{B}\Pi_t^{nonpert}=\int_{0}^{1}dx~\frac{1}
{144~\pi~M^6~x^4~(-1+x)^4} \exp\Big[\frac{m^2}{M^2~x~(-1+x)}\Big]
\Big\{\langle\alpha_{s}G^{2} \rangle \Big[12~M^6~x^4~
(-1+x)^4\nonumber\\&-&m^6~(1-2x)^2
(-1-x+x^2)-12~m^2~M^4~x^2~(-1+x)^2~(1-3x+3x^2)
+m^4~M^2~x~(-2+19x\nonumber\\&-&32x^2+11x^3+6x^4-2x^5)\Big]
+ 4~\alpha_{s}\langle \Theta^{g}\rangle\Big[
-8~M^6~x^3~(1-2x)^2(-1+x)^3+m^6~(1-2x)^2 (-1-x\nonumber\\&+&x^2)-2~m^2~M^4~x^2
~(-1+x)^2
(-1-6x+8x^2-4x^3+2x^4)+m^4~M^2~x~(-2+3x-12x^2+31x^3\nonumber\\&-&30x^4+10x^5)\Big]
  \Big\} ,
\end{eqnarray}
where, $\Theta^{g}=\Theta^{g}_{00}$.

\section{Numerical analysis}

In this section, we discuss the sensitivity of the masses and leptonic decay
constants of the  $J/\psi$ and $\Upsilon$ vector mesons to  temperature and obtain the numerical results for these quantities in vacuum. Taking into account  the  Eqs. (\ref{lim}) and
(\ref{BorelNonPert}) and applying derivative with respect to
$1/M^2$ to both sides of the  Eq. (\ref{f2m2}) and dividing by themselves, we obtain
\begin{eqnarray}\label{mV2}
m^2_{V}(T)=\frac{\int^{s_0(T)}_{4m^2}ds~s~
[\rho_{t,a}(s)+\rho_{\alpha_s}(s)]\exp\Big(-\frac{s}{M^2}\Big)+\Pi_1^{nonpert}(M^2,T)}
{\int^{s_0(T)}_{4m^2}ds
 [\rho_{t,a}(s)+\rho_{\alpha_s}(s)]\exp\Big(-\frac{s}{M^2}\Big)+\widehat{B}\Pi_t^{nonpert}},
\end{eqnarray}
where
\begin{eqnarray}\label{Pi1nonpert}
\Pi_1^{nonpert}(M^2,T)=M^4 \frac{d}{dM^2}\widehat{B}\Pi^{nonpert}_t,
\end{eqnarray}
and
\begin{eqnarray}\label{Pita}
\rho_{t,a}(s)=\frac{1}{8\pi^2}s\nu(s)(3-\nu^2(s))\left[1-2n\left(
\frac{\sqrt{s}}{2T}\right)\right] .
\end{eqnarray}
As we did also in \cite{E.V.Veliev3},  we  use the gluonic part
of the energy density both obtained from lattice QCD
 \cite{M.Cheng,D.E.Miller} and chiral perturbation theory \cite{P.Gerber}. In the rest
  frame of the heat bath, the results of some quantities obtained using
 lattice QCD in  \cite{M.Cheng}
 are well fitted by the help of the following  parametrization for the thermal average of total energy density $\langle \Theta \rangle$:
 \begin{eqnarray}\label{tetag}
\langle \Theta \rangle= 2 \langle \Theta^{g}\rangle=
6\times10^{-6}exp[80(T-0.1)](GeV^4),
\end{eqnarray}
where temperature $T $ is measured in units of $GeV$ and
this parametrization is valid only in the region   $0.1~GeV\leq T
\leq 0.17~GeV$. Here, we should stress that the total energy
density has been calculated for $T\geq 0$ in chiral perturbation
theory, while this quantity has only been obtained for $T\geq
100~MeV$  in lattice QCD (see
\cite{M.Cheng,D.E.Miller} for more details). In low temperature chiral perturbation
limit, the thermal average of the  energy density is expressed as
\cite{P.Gerber}:
\begin{eqnarray}\label{tetagchiral}
\langle \Theta\rangle= \langle \Theta^{\mu}_{\mu}\rangle +3~p,
\end{eqnarray}
where $\langle \Theta^{\mu}_{\mu}\rangle$ is trace of the
total energy momentum tensor and $p$ is pressure. These quantities
are given by:
\begin{eqnarray}\label{tetamumu}
\langle
\Theta^{\mu}_{\mu}\rangle&=&\frac{\pi^2}{270}\frac{T^{8}}{F_{\pi}^{4}}
\ln \Big(\frac{\Lambda_{p}}{T}\Big),\nonumber\\
p&=&
3T\Big(\frac{m_{\pi}~T}{2~\pi}\Big)^{\frac{3}{2}}\Big(1+\frac{15~T}{8~m_{\pi}}+\frac{105~T^{2}}{128~
m_{\pi}^{2}}\Big)exp\Big(-\frac{m_{\pi}}{T}\Big),\nonumber\\
\end{eqnarray}
where $\Lambda_{p}=0.275~GeV$, $F_{\pi}=0.093~GeV$ and
$m_{\pi}=0.14~GeV$.

The next step is to present the temperature dependent
continuum threshold and gluon
condensate. In the present work, we use the  $s_0(T)$  \cite{C.A.Dominguez2} and  $\langle G^2\rangle$  \cite{M.Cheng,D.E.Miller} as:
\begin{eqnarray}\label{sT}
s_0(T)= s_{0}\left[\vphantom{\int_0^{x_2}}
1-\Big(\frac{T}{T^{*}_{c}}\Big)^8\vphantom{\int_0^{x_2}}\right]+4~m_Q^2~
\left(\vphantom{\int_0^{x_2}}\frac{T}{T^{*}_{c}}
\vphantom{\int_0^{x_2}} \right)^8 ,
\end{eqnarray}
where $T^{*}_{c}=1.1~T_c=0.176~GeV$.
\begin{eqnarray}\label{G2TLattice}
\langle G^2\rangle=\frac{\langle
0|G^2|0\rangle}{exp\left[\vphantom{\int_0^{x_2}}12\Big(\frac{T}{T_{c}}-1.05\Big)
\vphantom{\int_0^{x_2}}\right]+1} .
\end{eqnarray}

\begin{figure}[h!]
\begin{center}
\includegraphics[width=8cm]{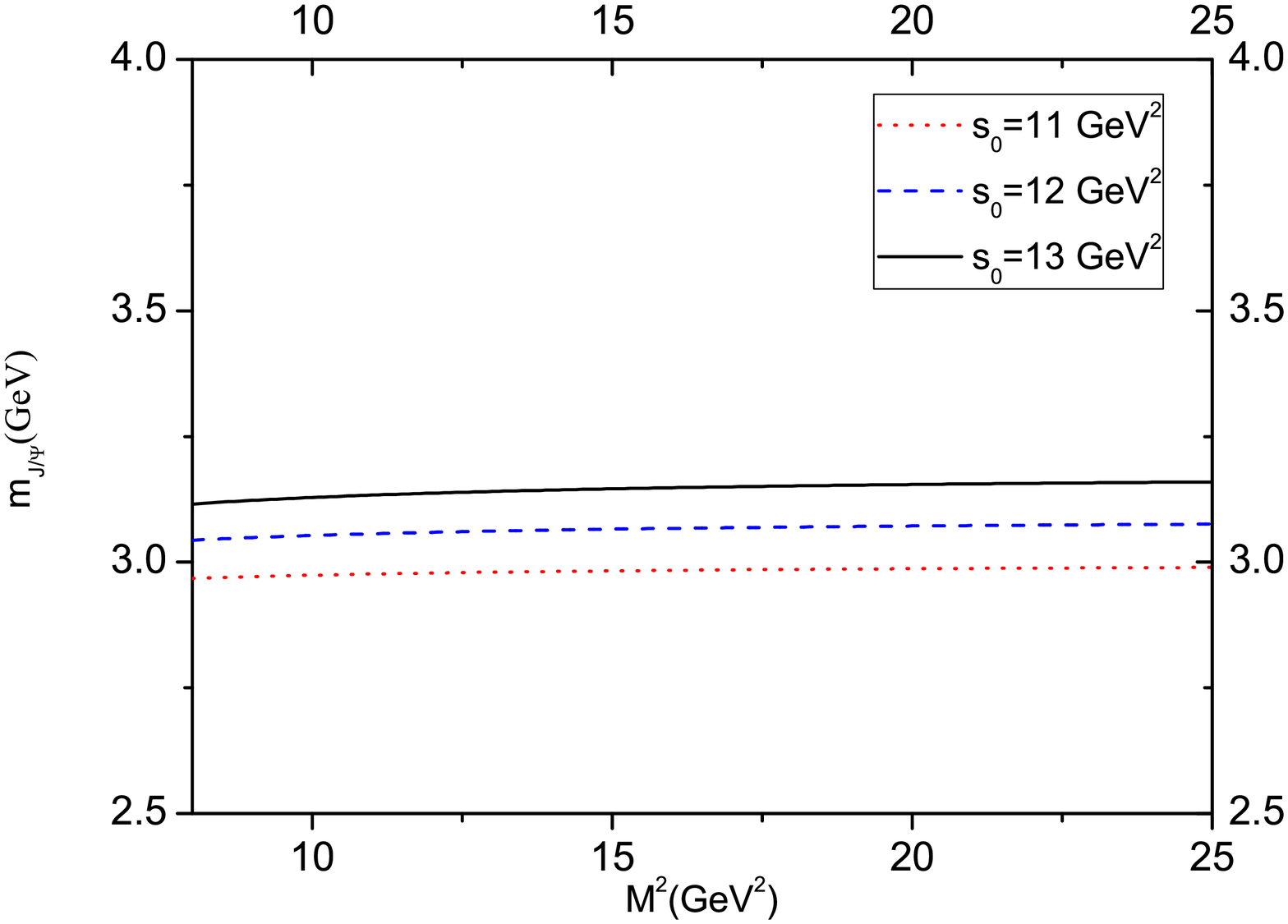}
\end{center}
\caption{The dependence of the mass of $J/\psi$  meson in vacuum on
the Borel parameter $M^2$.} \label{mJPsiMsq17Jan}
\end{figure}
\begin{figure}[h!]
\begin{center}
\includegraphics[width=8cm]{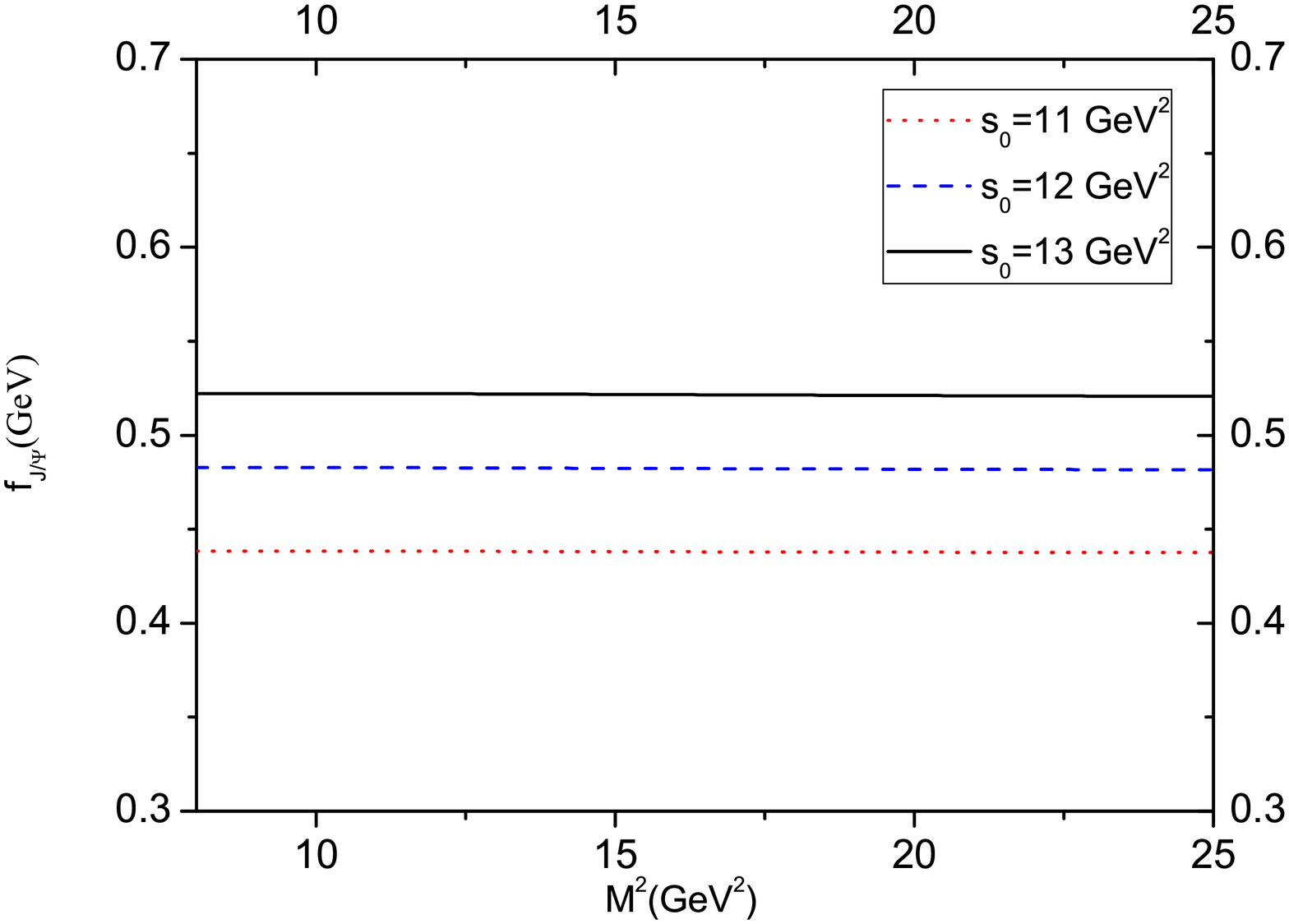}
\end{center}
\caption{The dependence of the leptonic decay constant of $J/\psi$
meson in vacuum on the Borel parameter $M^2$.} \label{fJPsiMsq17Jan}
\end{figure}

\begin{figure}[h!]
\begin{center}
\includegraphics[width=8cm]{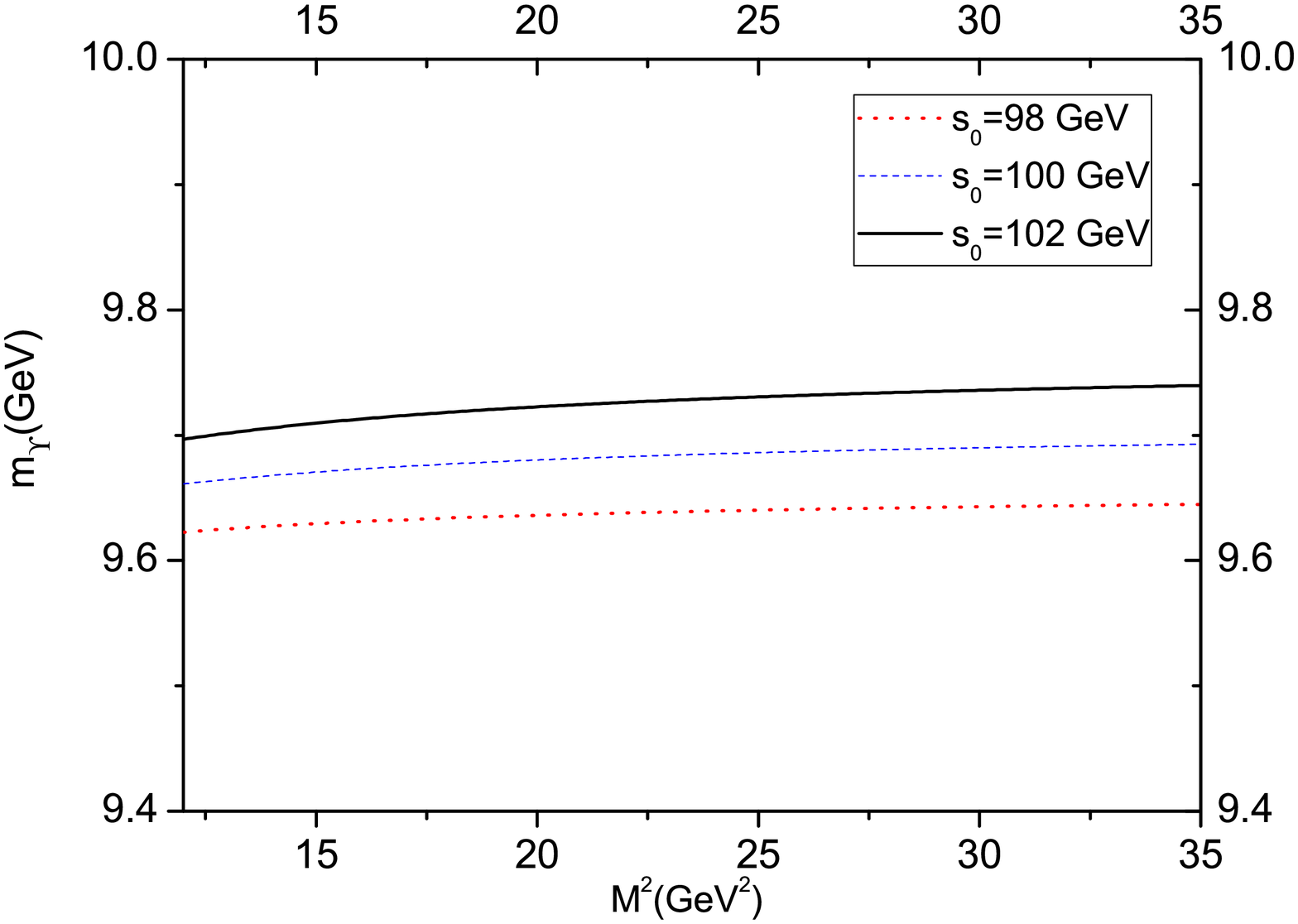}
\end{center}
\caption{The dependence of the mass of $\Upsilon$ meson in vacuum on the
Borel parameter $M^2$.} \label{mYMsq}
\end{figure}
\begin{figure}[h!]
\begin{center}
\includegraphics[width=8cm]{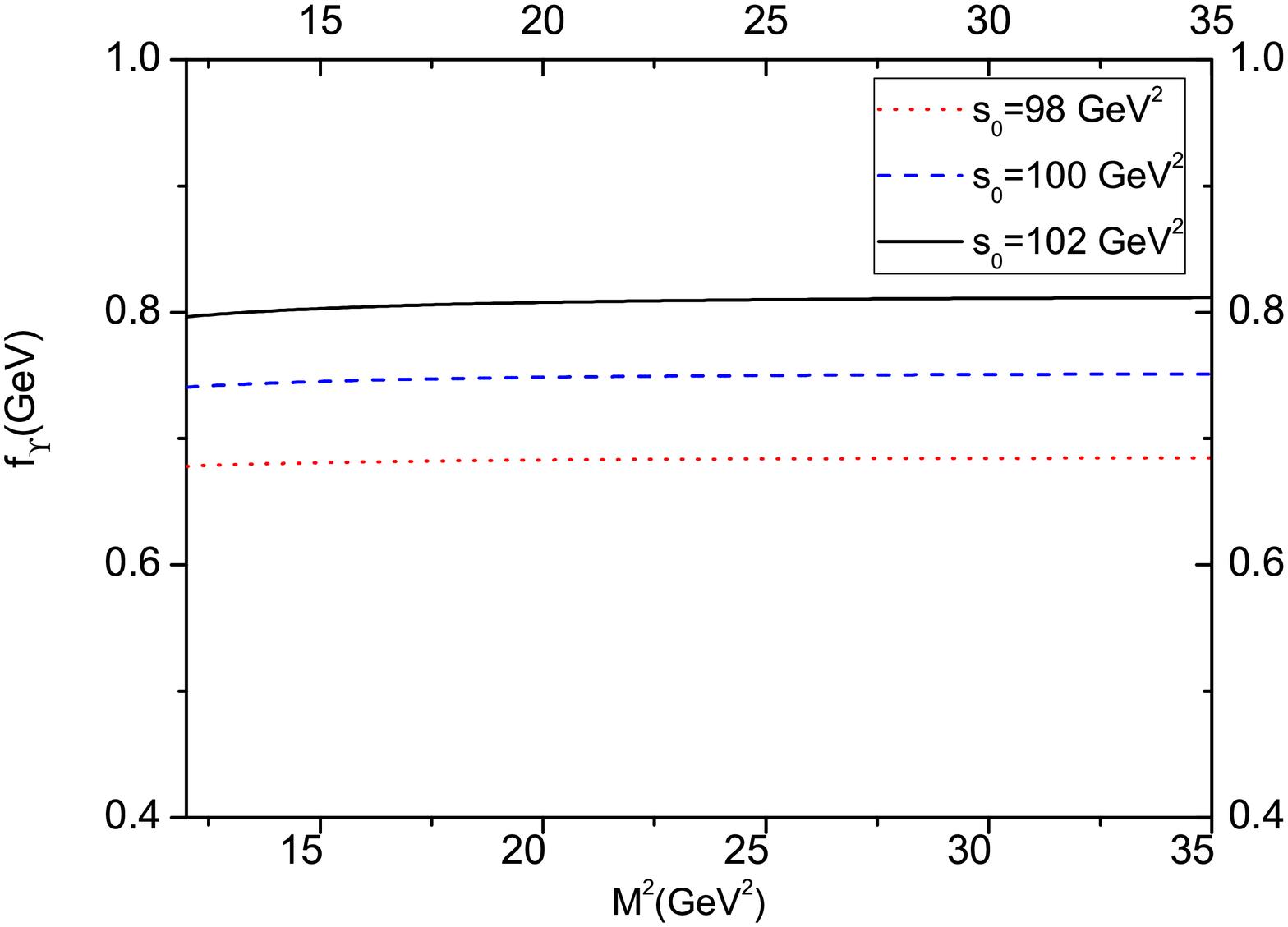}
\end{center}
\caption{The dependence of the leptonic decay constant of $\Upsilon$ meson
in vacuum on the Borel parameter $M^2$.} \label{fYMsq}
\end{figure}

In further analysis, we use the values, $m_c=(1.3\pm0.05)~GeV$,
$m_b=(4.7\pm0.1)~GeV$ and ${\langle}0\mid \frac{1}{\pi}\alpha_s G^2
\mid 0 {\rangle}=(0.012\pm0.004)~GeV^4$ for quarks
 masses and gluon condensate at zero temperature.
The  sum rules  for the masses and decay constants  also 
 include two more auxiliary parameters:  continuum
threshold $s_0$ and Borel mass parameter $M^2$.
These are not physical quantities, hence
 the physical observables should be approximately insensitive to these parameters. Therefore, we  look for working regions of these parameters
 such that the dependences of the masses and decay constants on these parameters are weak.  The continuum threshold, $s_{0}$ is not
completely arbitrary,  but it is  related to the energy of
the first exited state with the same quantum numbers as the
 interpolating currents. Our numerical analysis show that  in the intervals $s_0=(11- 13)~GeV^2$ and $s_0=(98- 102)~GeV^2$, respectively for   the  $J/\psi$ and $\Upsilon$ channels,
the results weakly depend on this parameter. The working region for the Borel mass parameter, $M^2$
is determined 
demanding that both the contributions of the higher states and continuum are sufficiently
suppressed and the contributions coming from the higher dimensional operators are small. As a result, the working region for the Borel parameter is found to be
$ 8~ GeV^2 \leq M^2 \leq 25~ GeV^2 $ and $ 12~ GeV^2 \leq M^2 \leq
35~ GeV^2 $ in  $J/\psi$ and $\Upsilon$ channels, respectively.

\begin{table}[h]
\renewcommand{\arraystretch}{1.5}
\addtolength{\arraycolsep}{3pt}
$$
\begin{array}{|c|c|c|c|}
\hline \hline
         &f_{J/\psi}(MeV) & f_{\Upsilon}(MeV)   \\
\hline
  \mbox{Present Work}        &  481\pm36   &  746\pm62 \\
\hline
  \mbox{Lattice\cite{V.V.Kiselev,O.Lakhina}}        & 399\pm4   &  - \\
\hline
  \mbox{Experimental \cite{V.V.Kiselev,O.Lakhina} }        & 409\pm15  &  708\pm8\\
\hline
  \mbox{Potential Model \cite{V.V.Kiselev}       } & 400\pm45  &  685\pm30\\
\hline
  \mbox{Nonrelativistic Quark Model \cite{O.Lakhina}   }     & 423  &  716\\
                    \hline \hline
\end{array}
$$
\caption{Values of the leptonic decay constants of the heavy-heavy
 $J/\psi$ and $\Upsilon$ vector  mesons in vacuum.} \label{tab:lepdecconst}
\renewcommand{\arraystretch}{1}
\addtolength{\arraycolsep}{-1.0pt}
\end{table}

\begin{table}[h]
\renewcommand{\arraystretch}{1.5}
\addtolength{\arraycolsep}{3pt}
$$
\begin{array}{|c|c|c|c|}
\hline \hline
 &  m_{J/\psi}~(GeV)& m_{\Upsilon}~(GeV)\\
\hline
  \mbox{Present Work }       &  3.05\pm0.08   &  9.68\pm0.25 \\
\hline
 \mbox{Experimental \cite{K. Nakamura}} &  3.096916\pm 0.000011 &9.46030\pm0.00026   \\
 \hline \hline
\end{array}
$$
\caption{Values of the masses of the heavy-heavy 
$J/\psi$ and $\Upsilon$ vector mesons in vacuum.} \label{tab:mass}
\renewcommand{\arraystretch}{1}
\addtolength{\arraycolsep}{-1.0pt}
\end{table}
\begin{figure}[h!]
\begin{center}
\includegraphics[width=10cm]{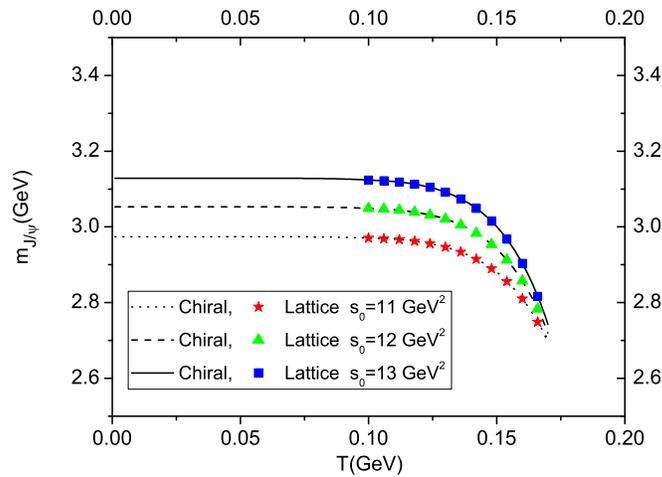}
\end{center}
\caption{The dependence of the mass of $J/\psi$ vector meson in GeV
on temperature at $M^2=10~GeV^2$.}
\label{mJPsiTempMsq10Last}
\end{figure}
\begin{figure}[h!]
\begin{center}
\includegraphics[width=10cm]{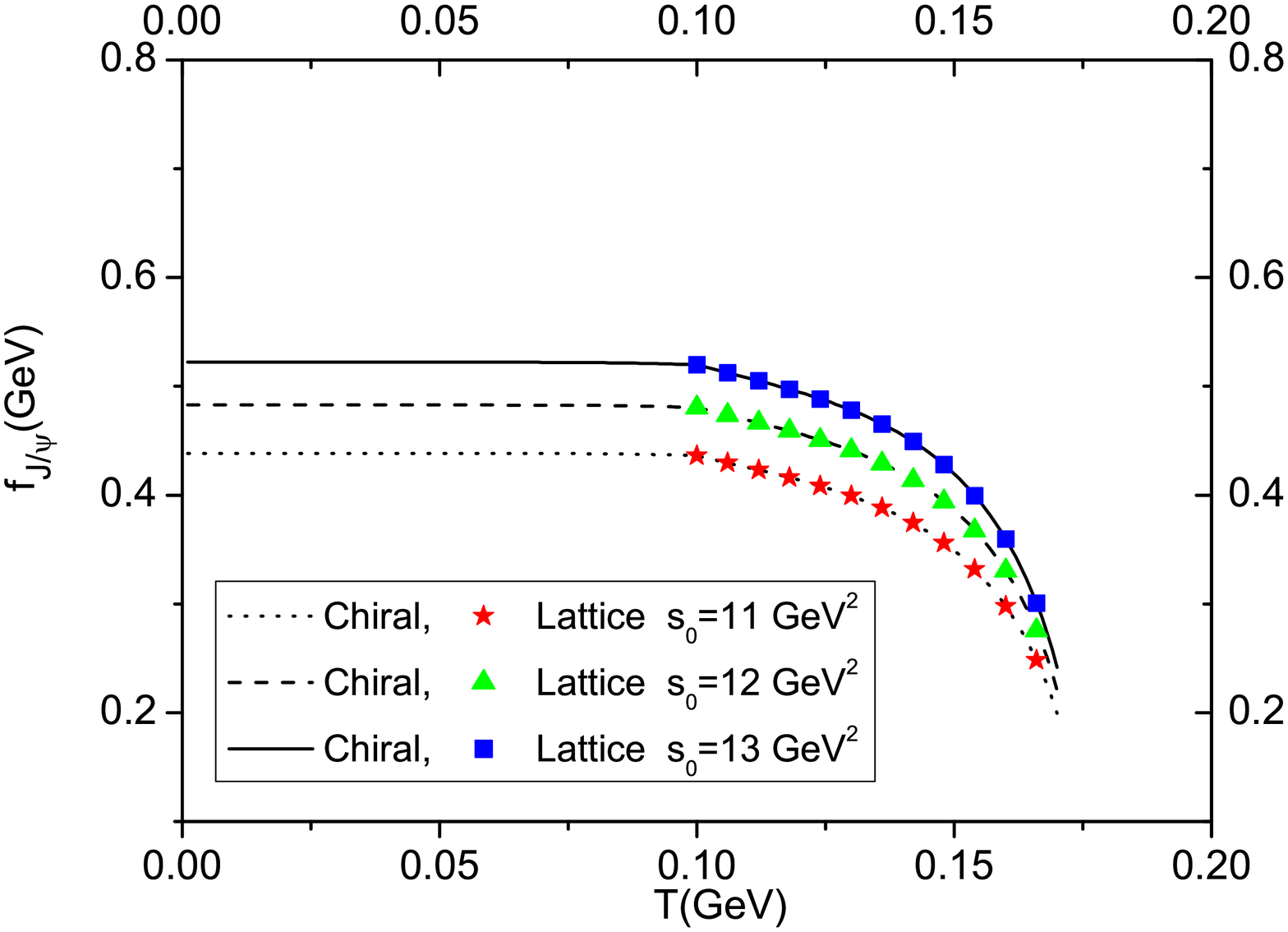}
\end{center}
\caption{The dependence of the leptonic decay constant of $J/\psi$
vector meson in GeV on temperature at $M^2=10~GeV^2$.} \label{fJPsiTempMsq10Last}
\end{figure}
\begin{figure}[h!]
\begin{center}
\includegraphics[width=10cm]{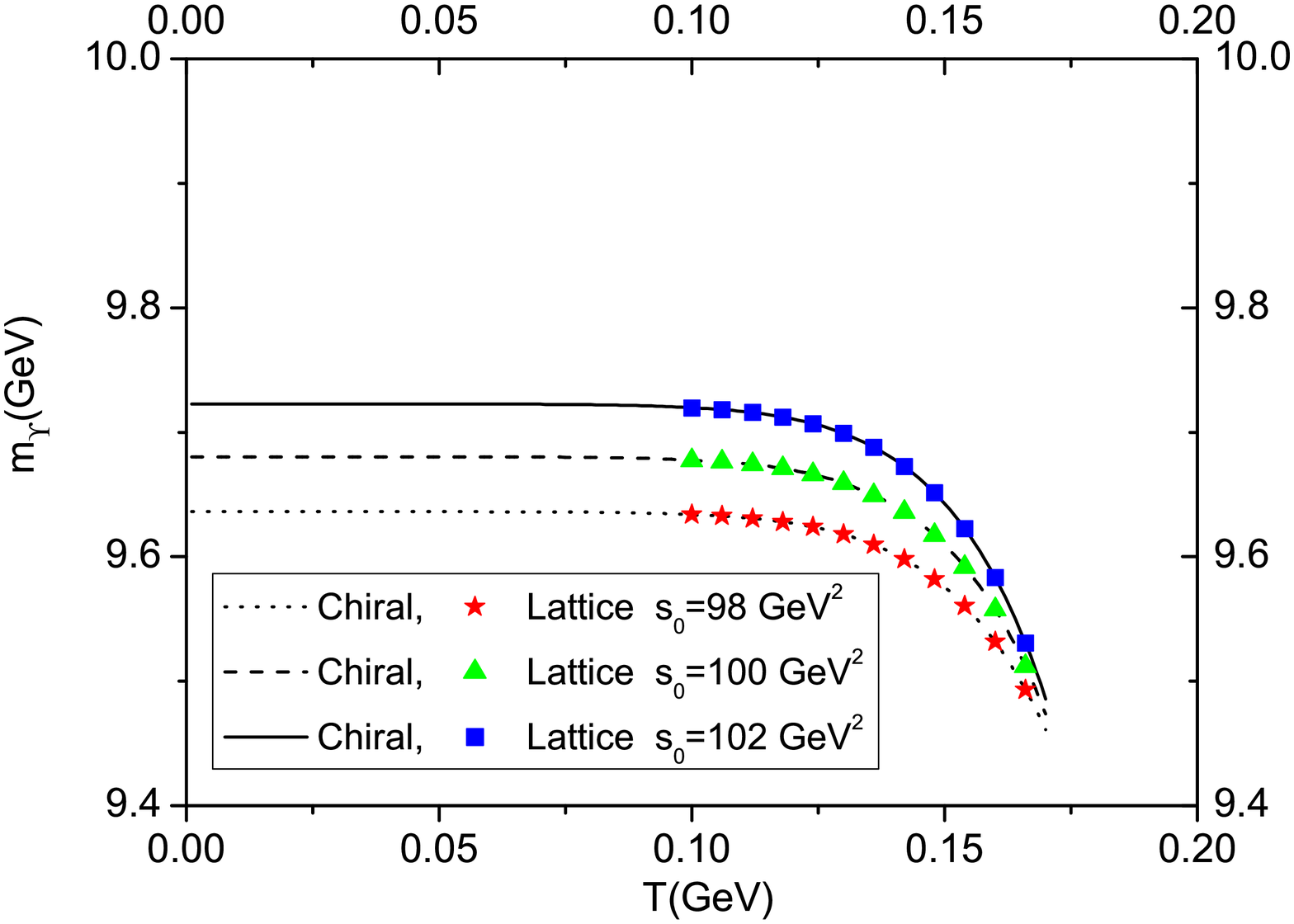}
\end{center}
\caption{The dependence of the mass of $\Upsilon$ vector meson in GeV on
temperature at $M^2=20~GeV^2$.}
\label{mYTempMsq20}
\end{figure}
\begin{figure}[h!]
\begin{center}
\includegraphics[width=10cm]{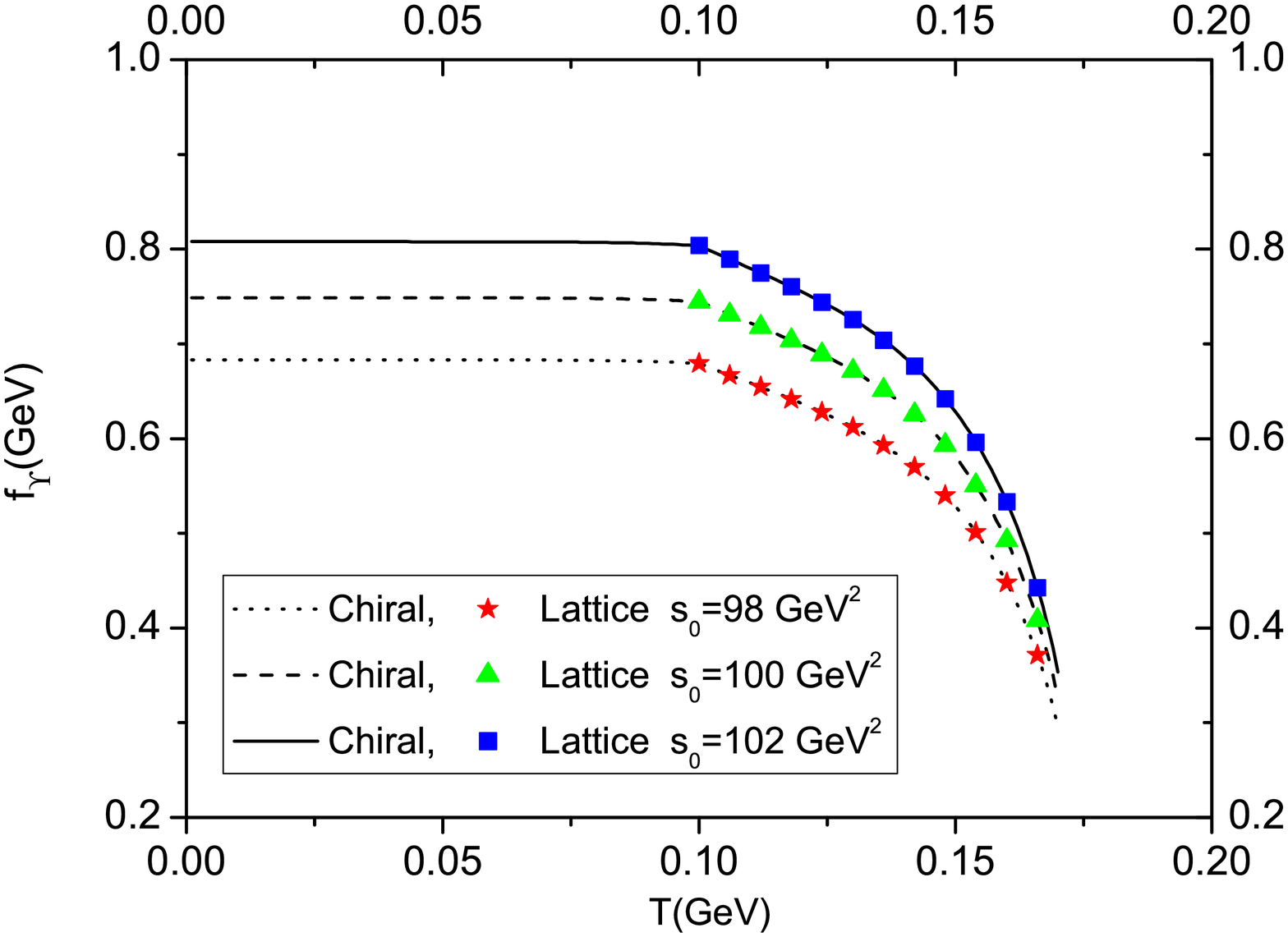}
\end{center}
\caption{The dependence of the leptonic decay constant of $\Upsilon$ vector
meson in GeV on temperature at $M^2=20~GeV^2$.} \label{fYTempMsq20}
\end{figure}

Using the working regions for auxiliary parameters as well as other input parameters, we depict the dependence of the masses and leptonic decay constants of  the heavy  $J/\psi$
and $\Upsilon$ vector quarkonia in Figs. (\ref{mJPsiMsq17Jan}-\ref{fYMsq}) at $T=0$ (vacuum). By a quick glance
in these figures, we see that the masses and decay constants represent good stability with respect to the variation of the Borel parameter in its working region. Also, we see a weak dependence of the results on
 the continuum threshold, $s_0$. From these figures, we deduce the numerical values of these parameters as shown in the Tables (\ref{tab:lepdecconst}) and (\ref{tab:mass}). The uncertainties presented in these Tables
are due to the uncertainties in calculation of the working regions for the auxiliary parameters as well as errors in the values of the other input parameters. In these Tables, we also present the 
existing numerical predictions from the other approaches such as lattice QCD, potential model and nonrelativistic quark model as well as existing experimental data. As far as the leptonic decay constants are concerned,
 our predictions for the central values are a bit bigger than the predictions of the other approaches and experiment, but when taking into account the uncertainties, our results become comparable especially with the 
potential   and nonrelativistic quark models predictions as well as experimental data. However, our predictions on the masses are in good consistency with the experimental values.

Our final task is to discuss the temperature dependence of the leptonic decay constant and masses of the considered particles. For this aim, we plot these quantities in terms of temperature in
 figures (\ref{mJPsiTempMsq10Last}-\ref{fYTempMsq20}) using the total energy density from both chiral perturbation theory
and lattice QCD (valid only for $T\geq
100~MeV$) and at different fixed values of the $s_0$ but a fixed value of the  Borel mass parameter.  From these figures, we observe    that the masses and decay
constants remain insensitive to the variation of the temperature up to $T\cong 100 ~MeV$, however after this
point, they start to diminish increasing the temperature.
At deconfinement or critical temperature, the decay constants approach
roughly to 45\% of their values at zero temperature, while the masses are
decreased about 12\%, and 2.5\%  for $J/\psi$ and $\Upsilon$ states,
respectively. Considerable decreasing in the values of the decay constants near the deconfinement temperature can be judged as a sign of the quark gluon plasma  phase transition.

\section{Acknowledgement}

The authors are grateful to
 T. M. Aliev for useful
discussions. This work is supported in part by the Scientific and
Technological Research Council of Turkey (TUBITAK) under the
research project No. 110T284.


\begin{thebibliography}{99}
%
%
%
\bibitem{Shifman} M. A. Shifman, A. I. Vainstein,  V. I. Zakharov,  Nucl. Phys. \textit{B147}, 385 (1979);
M. A. Shifman,  A. I.  Vainstein, V. I. Zakharov,  Nucl. Phys.
\textit{B147}, 448 (1979).
%
\bibitem{Matsui and Satz} T. Matsui, H. Satz, Phys. Lett. \textit{B178}, 416 (1986).
%
\bibitem{L.J.Reinders} L. J. Reinders, H. Rubinstein and S. Yazaki,  Phys. Rep. \textit{127},
No1 (1985) 1.
%
\bibitem{P. Colangelo} P. Colangelo, A. Khodjamirian, in At the Frontier of Particle
Physics/Handbook of QCD, edited by M. Shifman (World Scientific,
Singapore, 2001), Vol. 3, p. 1495.
%
\bibitem{Bochkarev} A. I. Bochkarev,  M. E. Shaposhnikov,  Nucl. Phys.
\textit{B268}, 220, (1986).
%
\bibitem{E.V. Shuryak} E.V. Shuryak, Rev. Mod. Phys. 65, 1 (1993).
%
\bibitem{T. Hatsuda} T. Hatsuda, Y. Koike, S.H. Lee, Nucl. Phys. \textit{B394}, 221 (1993).
%
\bibitem{S. Mallik} S. Mallik, Phys. Lett. \textit{B416}, 373 (1998).
%
\bibitem{S.Mallik Mukherjee} S. Mallik, K. Mukherjee, Phys. Rev. \textit{D58},
096011 (1998); Phys. Rev. \textit{D61}, 116007 (2000).
%
\bibitem{S.Mallik sarkar} S. Mallik, S. Sarkar, Phys.Rev.
\textit{D66}, 056008 (2002).
%
\bibitem{E.V.Veliev} E. V. Veliev, J. Phys. G:Nucl. Part. Phys., \textit{G35}, 035004 (2008);
E. V. Veliev, T. M. Aliev, J. Phys. G:Nucl. Part. Phys., \textit{G35}, 125002 (2008).
%
\bibitem{C.A.Dominguez} C.A. Dominguez, M. Loewe, J.C. Rojas, JHEP 08, 040 (2007).
%
\bibitem{C.A.Dominguez2} C. A. Dominguez, M. Loewe, J.C. Rojas, Y. Zhang,   Phys. Rev. \textit{D81},
 014007 (2010).

\bibitem{E.V.Veliev2} E. V. Veliev, G. Kaya, Eur. Phys. J.
\textit{C63}, 87 (2009); Acta Phys. Polon. \textit{B41}, 1905 
(2010).
%
\bibitem{F. Klingl} F. Klingl, S. Kim, S.H. Lee, P. Morath and W. Weise, Phys. Rev.
Lett. 82, (1999).
%
\bibitem{K.Morita} K. Morita, S.H. Lee, Phys. Rev.
\textit{C77}, 064904 (2008).
%
\bibitem{K.Morita2} K. Morita, S.H. Lee,Phys. Rev.
 \textit{D82}, 054008 (2010).
%
\bibitem{E.V.Veliev3} E. V. Veliev, H. Sundu, K. Azizi, M. Bayar,  Phys. Rev.
\textit{D82}, 056012 (2010); E. V. Veliev, K. Azizi, H. Sundu, N.
Ak\c sit, arXiv:1010.3110 [hep-ph].
%
\bibitem{P.Gerber} P. Gerber, H. Leutwyler, Nucl. Phys.
\textit{B321}, 387 (1989).
%
\bibitem{M.Cheng} M. Cheng, et.al, Phys. Rev.
\textit{D77}, 014511 (2008).
%
\bibitem{D.E.Miller} D. E. Miller, Phys. Rept.
\textit{443}, 55 (2007).
%
\bibitem{O.Kaczmarek} O. Kaczmarek, F. Karsch, F. Zantow, P. Petreczky, Phys. Rev.
\textit{D70}, 074505 (2004).
%
\bibitem{R.L.Kobes} R. L. Kobes, G.W. Semenoff, Nucl. Phys. \textit{260}, 714 (1985), S.
Sarkar, B. K. Patra, V. J. Menon,  S. Mallik, Indian J. Phys.
\textit{76A}, 385 (2002)
%
\bibitem{A.Das} A. Das, Finite Temperature Field Theory, World Scientific
(1999).
%
\bibitem{V.V.Kiselev} V. V. Kiselev, A. K. Likhoded, O. N. Pakhomova, V. A. Saleev, Phys. Rev.
\textit{D65}, 034013(2002).
%
\bibitem{O.Lakhina} O. Lakhina, E. S. Swanson, Phys. Rev.
\textit{D74}, 014012 (2006).
%
\bibitem{K. Nakamura}  K. Nakamura et al. (Particle Data Group),
J. Phys. \textit{G37}, 075021 (2010).
%
%
%


\end{thebibliography}
\end{document}